\documentclass[doublecol]{epl2} 
\usepackage{graphicx}
\usepackage{amssymb}
\usepackage{amsmath}

\title{
Pattern orientation in finite domains  without boundaries
}
\shorttitle{Orientational transition of patterns} 

\author{Lisa Rapp \and Fabian Bergmann \and Walter Zimmermann}
\shortauthor{L. Rapp \etal}

\institute{Theoretische Physik I, Universit\"at Bayreuth, 95440 Bayreuth, Germany}

\abstract{
We investigate the orientation of nonlinear stripe patterns in finite domains. 
Motivated by recent experiments, we introduce a control parameter drop
from supercritical inside a domain to subcritical outside without boundary conditions at the domain border. 
As a result, stripes align perpendicular to shallow control parameter drops. 
For steeper drops, non-adiabatic effects lead to a surprising orientational transition to parallel stripes with respect to the borders. 
We demonstrate this effect in terms of the Brusselator model and generic amplitude equations.
}

\begin{document}

\maketitle

\section{Introduction}
Pattern formation is central to the wealth of fascinating phenomena in nature.
It occurs in a great variety of physical, chemical and living systems \cite{Ball:98,Cross:2009}. 
Examples include patterns in isotropic and anisotropic convection systems \cite{CrossHo,Lappa:2010, Weiss:2012, Zimmermann:88.3,Kramer:95.1}, chemical reactions \cite{Kapral:1995,Mikhailov:2006.1} 
and biological systems \cite{BenJacob:1994.1,Kondo:2010.1,Schwille:2008.1}, or environmental patterns \cite{Meron:2015}.

In real systems, patterns emerge in  finite areas or volumes. 
Consequently, spatially periodic patterns  only contain a finite number of wavelengths. 
Along the system borders, the relevant fields have to obey boundary conditions that influence the pattern in different ways \cite{Greenside:84.1,CrossHo,Cross:1982.2,Ahlers:1999.1,Greenside:2003.1,CDHS,Cross:1986.1,Cross:1988.1,Steinberg:87.1,Kolodner:88.1,Daviaud:2003.2}.
In isotropic systems, they render the orientation of stationary patterns \cite{Greenside:84.1,CrossHo}.
In thermal convection, convection rolls align perpendicular to side walls 
due to boundary conditions for the flow fields \cite{Cross:1982.2,Ahlers:1999.1,Greenside:2003.1}. 
Boundary conditions at the side walls may also restrict the range of possible stable wave numbers of periodic patterns \cite{CDHS}.
Traveling waves of finite wave number may be reflected at the boundaries leading
to a number of interesting and complex phenomena \cite{Cross:1986.1,Cross:1988.1,Steinberg:87.1,Kolodner:88.1,Daviaud:2003.2}.

\begin{figure}
\begin{center}
\includegraphics[width=\columnwidth]{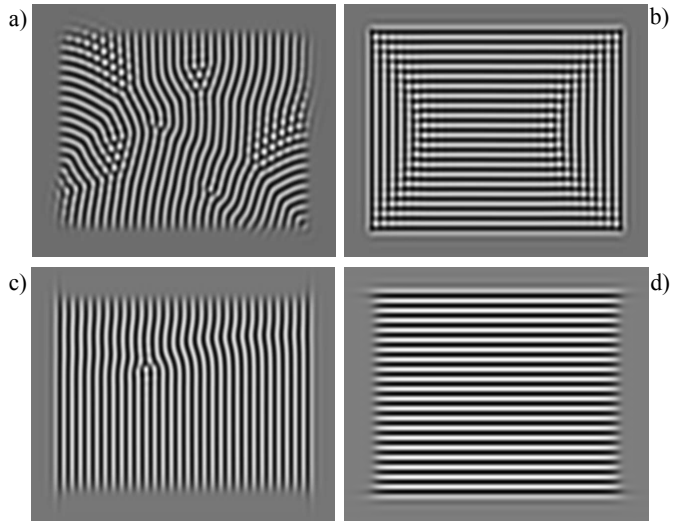}
\end{center}
\vspace{-2mm}
\caption{Stripe patterns inside supercritical subdomains in the Brusselator model.
The control parameter drops 
on different length scales
$\delta_{x,y}$ along $x$ and $y$ 
from $\beta_m=0.05$ to subcritical values in a wide
vicinity:
a) $\delta_x=\delta_y=\lambda_c$, b) $\delta_x=\delta_y=0.32\lambda_c$, c) $\delta_x=0.32\lambda_c$, $\delta_y=1.5\lambda_c$,
d) $\delta_x=1.5\lambda_c$, $\delta_y=0.32\lambda_c$.}
\vspace{-5mm}
\label{Abb_rectangularboxes}
\end{figure}

However, finite systems can also be achieved when the fluxes and forces driving a pattern, 
the so-called control parameters, are sufficiently strong (supercritical) only in a subdomain of the system.
In this case, no specific boundary conditions act on the fields at {\it control parameter drops} to subcritical values. 
The effects of restricting patterns to a finite domain  in this way have not been systematically investigated before. 
Examples of pattern orientations resulting from different widths of the control parameter drops 
are shown in fig.~\ref{Abb_rectangularboxes} and explained in this work.

Recent  experiments where pattern forming  protein reactions take place 
in finite subdomains of substrates \cite{Schwille:2012.1} belong to this class. 
Control parameter drops can also be designed in light sensitive chemical reactions 
where illumination of the reaction cell suppresses pattern formation  \cite{Epstein:1999.1,Epstein:2001.1}. 
If the illumination is only applied to a subdomain of the system, again no boundary conditions for the concentration fields are defined along the edge of the illumination mask.

We investigate how control parameter drops along the borders of a supercritical subdomain affect the orientation of  stationary spatially periodic patterns
when no boundary conditions for the fields are specified.
We choose the Brusselator as a representative model system to study the influence of the control parameter drop width.
This is complemented by studies of the so-called amplitude equations
for supercritical bifurcations to spatially periodic patterns \cite{CrossHo}. 
As a general description for this class of patterns, the conclusions drawn from the amplitude equations emphasize the universality of our results.

For large drop widths, we find that stripes align perpendicular to the borders of
the supercritical control parameter domain.
By decreasing the length scale for the control parameter drop,
we find a surprising  orientational  transition to stripes in parallel alignment. 
The analysis of the amplitude equations reveals additional non-adiabatic, resonance-like effects favouring parallel stripes.

\section{Model systems and control parameter drop}
%

\subsection{Brusselator}
The Brusselator  is a common model for reaction-diffusion systems \cite{Prigogine:1969.1,Nicolis:77, Walgraef:1997, PerezG:2001.1}.
We use it as a prototype system for supercritical bifurcations to spatially periodic 
patterns (Turing patterns). 
It describes the nonlinear behaviour of the concentration fields $u(x,y,t)$ and $v(x,y,t)$:
\begin{subequations}
\label{Brusselator}
\begin{align}
\partial_t u&=\nabla^2 u+a-(b+1)u+u^2v\,,\label{DglUIso}\\
\partial_t v&=D\nabla^2v+bu-u^2v \label{DglVIso},
\end{align}
\end{subequations}
with the {\it control parameter} $b$ and constant parameters $a$, $D$.
These equations have the homogeneous fixed point solution
\begin{align}
\label{Brusshomfix}
 u_h=a\,, \qquad v_h=b/a\,.
\end{align}
Turing patterns with the critical wave number $q_c$ bifurcate from this basic state for control parameter 
values beyond its critical one  $b_c$ \cite{Walgraef:1997}, where
\begin{align}
\label{Brussbasic}
 b_c = (1+a\eta)^2\,, \qquad q_c=\sqrt{a\eta}\,,
\end{align}
and ${\eta :=\sqrt{1/D}}$. 
The relative distance $\beta$ of the control parameter from its critical value $b_c$ is 
given by
\begin{align}
\label{contbeta}
 b= b_c(1+\beta)\,,
\end{align}
\textit{i.e.} $\beta_c=0$.
Hexagons are typical for the Brusselator near the onset of Turing patterns. 
But in this work, we consider the special case ${D=a^2}$ where stripes are preferred at the onset \cite{PerezG:2001.1}.
In this case, the critical wavelength of the stripes according to eq.~(\ref{Brussbasic})
is $\lambda_c:=2\pi/q_c=2\pi$. We choose $a=4$ throughout this work. 
 
\subsection{Amplitude equations} 

The two concentration fields $u$ and $v$  may be combined to the vector field
${\bf w}({\bf r},t)=(u({\bf r},t), v({\bf r},t))$. 
We write spatially periodic stripes with the wave vector ${\bf q}_c$ in the form \cite{CrossHo,Walgraef:1997}
\begin{align}
\label{ampexp}
{\bf w}({\bf r},t)= {\bf w}_h + A \tilde {\bf w} \text{e}^{\text{i}({\bf q}_c \cdot {\bf r})}+A^\ast \tilde {\bf w}^\ast \text{e}^{-\text{i}({\bf q}_c \cdot {\bf r})}\,,
\end{align}
where ${\bf w}_h=(u_h,v_h)$.
Slow variations (compared to the wavelength $\lambda_c$) of the envelope $A({\bf r},t)$ can be described by a dynamical amplitude equation \cite{Newell:1969.1,CrossHo}.

The Brusselator model is isotropic. Hence, in extended systems only the magnitude $q_c$ of the critical wave vector ${\bf q}_c$ for Turing stripes is fixed, but not its direction. 
Thus, all stripe orientations are equally likely at pattern onset. 
We consider the amplitude equations in two limits of stripe orientations: 
 ${\bf q}_c=(q_c,0)$ and ${\bf q}_c=(0,q_c)$, called parallel and perpendicular hereafter.
The reduction method to amplitude equations, the so-called multiple scale analysis, is well established for supercritical bifurcations \cite{Newell:1969.1,CrossHo}. 
The generic amplitude equations for the two stripe orientations in the case of a small and constant control parameter $\beta$ are 
\begin{align}
 \label{Ampli_A}
 \partial_t A &=  \beta A  + {\cal L} A - g|A|^2A,
\end{align}
with
\begin{subequations}
\begin{align}
 {\cal L}&= {\cal L}_\parallel^2 := \xi_0^2\left(\partial_x -\frac{\text{i}}{2q_c} \partial_y^2\right)^2  \quad  \mbox{for} \quad {\bf q}_c=(q_c,0),\label{linop1} \\
{\cal L}&= {\cal L}_\perp^2 := \xi_0^2\left(\partial_y -\frac{\text{i}}{2q_c} \partial_x^2\right)^2 \quad \mbox{for} \quad {\bf q}_c=(0,q_c). \label{linop2}
 \end{align}
\end{subequations}
The coherence length $\xi_0$ and the
nonlinear coefficient $g$ for the Brusselator in the special case of ${D=a^2}$ are
$\xi_0^2=1$ and $g =3/(2a^2)$ \cite{PerezG:2001.1}.

\subsection{Control parameter drop}
We introduce the control parameter drop by assuming the spatially dependent control parameter $\beta(x,\delta_x)$:
\begin{align}
\beta=\beta_0+\frac{M}{2} \left[\tanh\left(\frac{x-x_l}{\delta_x}\right)
-\tanh\left(\frac{x-x_r}{\delta_x}\right) \right]\,.
\label{eq_Epsilon}
\end{align}
We assume $L:=x_r-x_l\gg\lambda_c$ and $\beta_0<0$. 
$M$ and $\beta_0$ are chosen such that the maximum value $\beta_m=\beta_0+M$ is small and positive. 
Then $\beta(x,\delta_x)$ is supercritical  in the  subdomain $\bar x_l<x < \bar x_r$, where
\begin{align}
\bar x_{l,r}=x_{l,r} \pm \frac{\delta_x}{2}~ \ln\left(\frac{-\beta_0}{M+\beta_0}\right)\,,
\end{align}
and drops down to the subcritical value $\beta_0$ outside this domain.
The steepness of the control parameter drop around $\bar x_{l,r}$ increases with decreasing values of the drop width  $\delta_x$.

For small values of $\delta_x$, the control parameter $\beta(x,\delta_x)$ 
varies rapidly in a narrow  range around $\bar x_{l,r}$. 
However, only the slowly (adiabatically) varying contributions to $\beta(x,\delta_x)$
affect the solutions of  amplitude equations.  
The rapidly (non-adiabatically) varying part is smoothed out and must be treated separately.
We therefore decompose $\beta(x,\delta_x)$ into an adiabatic and non-adiabatic part. 
For this purpose, we introduce the slow length scale $\delta_A:=2\xi_0/\sqrt{\beta_m}>\delta_x$ and choose $\beta_0=-\varepsilon$, $M=2\varepsilon$ (where $\varepsilon$ is positive and small).
We then express the slowly varying contribution $B_0(x)$ via eq.~(\ref{eq_Epsilon}) by choosing $\delta_x=\delta_A$:
\begin{equation}
\label{beta_adiabatic}
B_0(x)=\beta(x, \delta_A).
\end{equation}

\begin{figure}
\includegraphics[width=\columnwidth]{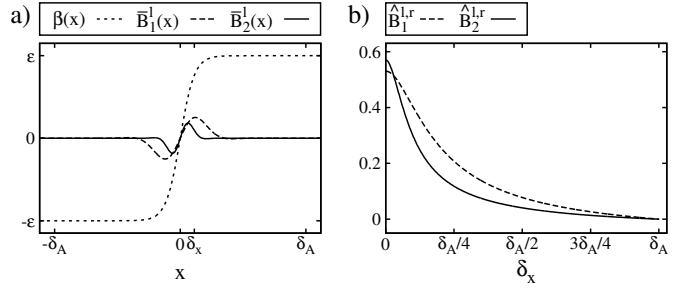}
\caption{a) Contributions $\bar B_1^l(x)$ and $\bar B_2^l(x)$ to the control parameter drop $\beta(x,\delta_x)$ 
for $\delta_x=0.11\delta_A$. b) Gaussian amplitudes $\hat{B}_1^{l,r}$ and $\hat{B}_2^{l,r}$ of the localised
amplitudes as a function of the drop width $\delta_x$ for $\varepsilon=0.05$.
\label{Abb_ResonantEpsPart}}
\end{figure}

The difference between $\beta(x, \delta_x)$ and $B_0(x)$  becomes small in the centre of $[ x_l ,  x_r]$ and takes its largest values around $x_{l,r}$.
We expand the rapidly varying difference $\beta(x, \delta_x)-B_0(x)$ into a series to obtain
\begin{align}
\label{Eps_decomp}
\beta(x, \delta_x)=B_0(x) &+\frac{M}{2} \sum_{m} \, \Bigl\{ B_{m}^l(x) \sin\left[m q_c(x- x_l)\right]  \nonumber \\   
& +B_{m}^r(x) \sin\left[m q_c(x-x_r)\right]\Bigr\},
\end{align}
where $m=n/N_L$, $N_L=L/\lambda_c$. 
The integer $n$ is chosen such that $m$ belongs to the range $1/8,1/7 ...4$.
The functions $B_m^{l,r}(x)$ are localised around $x_{l,r}$ and represented by Gaussians of the form, 
\begin{equation}
B_m^{l,r}(x)=\hat{B}_m^{l,r}\,\exp\left[-\frac{(x-x_{l,r})^2}{\delta_{G,m}^2} \right].
\end{equation}
The Gaussian amplitudes $\hat{B}_m^{l,r}$ and their widths $\delta_{G,m}$ are determined via a correlation analysis. 
We calculate the correlation function between the rapidly varying part 
\begin{equation}
\Delta\tilde{\beta}(x,\delta)=\tanh(x/\delta)-\tanh(x/\delta_A)
\end{equation}
and the test function
\begin{equation}
f_m(x,\delta_\text{test})=\frac{1}{\sqrt{\pi}\delta_\text{test}}\,\text{e}^{-x^2/\delta_\text{test}^2}\sin(m q_c x).
\end{equation}
We then choose the Gaussian width $\delta_{G,m}$ to be the value of $\delta_{test}$ that maximises the correlation function.
The amplitudes $\hat{B}_m^{l,r}$ are calculated via the overlap integral between $f_m(x, \delta_{G,m})$ and $\Delta\tilde{\beta}$.
Figure ~\ref{Abb_ResonantEpsPart}a) shows the contributions $\bar B_m^{l}:=\varepsilon B_m^l(x)\sin(mq_cx)$ for $m=1,2$ in comparison to the full shape of $\beta(x,\delta_x)$.
Both functions are localised around $x_l=0$ and approach zero within a short range ($\ll\delta_A$) around the control parameter drop. 
The Gaussian amplitudes $\hat{B}_1^{l,r}$ and $\hat{B}_2^{l,r}$ decrease
as a function of the drop width $\delta_x$ [fig.~\ref{Abb_ResonantEpsPart}b)]. 
These non-adiabatic contributions vanish for $\delta_x>\delta_A$. 
The amplitude $\hat{B}_1^{l,r}$ is usually larger than $\hat{B}_2^{l,r}$, except in the limit of very small drop widths. 
 
The patterns in fig.~\ref{Abb_rectangularboxes} are obtained for a rectangular supercritical subdomain of the control parameter in the form
\begin{align}
 \beta=\beta_0\,+\,& \frac{M}{4}\left[\tanh\left(\frac{x-x_l}{\delta_x}\right)
-\tanh\left(\frac{x-x_r}{\delta_x}\right) \right] \nonumber \\
& \times\,\,\left[\tanh\left(\frac{y-y_b}{\delta_y}\right)
-\tanh\left(\frac{y-y_t}{\delta_y}\right)\, \right]\,. 
\label{controldrop_2d}
\end{align}
Here, we introduced a second drop width $\delta_y$ to describe the additional spatial dependence of $\beta$ in the $y$-direction.
$\beta(x,y,\delta_x,\delta_y)$ is supercritical in the two-dimensional area $[\bar x_l,\bar x_r]\times[\bar y_b,\bar y_t]$.

\section{
Non-adiabatic effects cause an orientational transition}
\label{Envres}

We now include the control parameter drop into the amplitude equation using the decomposition given in eq.~(\ref{Eps_decomp}). 
The control parameter $\beta$ in eq.~(\ref{Ampli_A}) 
is replaced by the slowly (adiabatically) varying part $B_0(x)$ as given by
eq.~(\ref{beta_adiabatic}). 
The short wavelength contributions $\propto B_{m}^{l,r}(x) \exp{(\text{i}mq_c x)}$ ($m=1,2,3,4$) in eq.~(\ref{Eps_decomp}) cause additional (non-adiabatic) terms in the amplitude equation for parallel stripes \cite{Coullet:86.2}.  
It then takes the form
 \begin{align}
 \label{Ampli_A_resonance}
 \partial_t A = & B_0(x) A  + {\cal L}_\parallel^2 A - g|A|^2A \nonumber \\
 & + \sum_{m=1}^4 \alpha_m B_m(x) \left(A^\ast\right)^{m-1}.
\end{align}
Here, $\alpha_m$ are constant parameters depending on the respective system. 
The complex localised contributions $B_m(x)$ due to the control parameter drop are given by 
\begin{align}
B_m(x)=\text{i}\frac{M}{4}\left[B_m^l(x)\text{e}^{-\text{i}mq_c x_l}-B_m^r(x)\text{e}^{-\text{i}mq_c x_r}\right].
\end{align}
The effects caused by $B_{3,4}(x)$ are much smaller than $B_{1,2}(x)$ and therefore neglected in the following.
Equation (\ref{Ampli_A_resonance}) can be derived from the functional
\begin{align}
\label{Ampli_A_resonance_functional}
 F_\parallel &= \int dx dy\biggl[- B_0(x)|A|^2 + \frac{g}{2}|A|^4 + \left|{\cal L}_\parallel A\right|^2 
 \nonumber \\
  & \qquad - \sum_{m=1}^2 \frac{\alpha_m}{m}\left(B_m(x)A^{*^m}+B_m^{*}(x) A^m\right)  \biggr]
\end{align}
via $\partial_t A=-\delta F_\parallel/\delta A^{*}$. 
For the Brusselator in the case $D=a^2$, we find $\alpha_1=2a$ and $\alpha_2=5/3$.

The  amplitude equation for perpendicular stripes with ${\bf q}_c=(0,q_c)$ is not affected by 
resonance contributions $\propto B_{m}$. It is described by eq.~(\ref{Ampli_A}) with ${\cal L}={\cal L}_\perp^2$ as 
given in eq.~(\ref{linop2}) and the slowly varying control parameter $\beta=B_0(x)$, cf. eq.~(\ref{beta_adiabatic}). 
The related functional is
\begin{equation}
\label{Ampli_A_nonres_functional}
 F_\perp = \int dx dy\left[-B_0(x)|A|^2 + \frac{g}{2}|A|^4 
 + \left|{\cal L}_\perp A\right|^2\,\right].
\end{equation}

Considering small values of $\delta_x$, non-adiabatic effects described by $B_{1,2}$ play an important role. 
However, these only affect the amplitude equation for parallel stripes, cf. eq.~(\ref{Ampli_A_resonance}). 
A finite value of $B_1$ changes the supercritical pitchfork bifurcation (in the case $B_1=0$) to an imperfect bifurcation 
\cite{Coullet:86.2,PeterR:2005.1}.
Consequently, parallel stripes already have a finite amplitude below the bulk threshold $\beta_m=0$, especially around $x_{l,r}$, 
where $B_{1,2}$ take the largest values. 
This finite amplitude $A$ decreases the functional $F_{\parallel}$ for parallel stripes with respect to $F_{\perp}$. 
Thus, for small values of $\delta_x$, parallel stripes are preferred compared to perpendicular stripes.

For large values of $\delta_x$, the non-adiabatic contributions $B_{1,2}$ become small and can be 
neglected (see fig.~\ref{Abb_ResonantEpsPart}). In this case, the amplitude equations and the 
functionals for the two different stripe orientations only differ in the linear operator. 
These include different orders of derivatives in $x$-direction: 
$|\partial_x A|^2$ in the functional for parallel stripes, eq.~(\ref{Ampli_A_resonance_functional}), 
and $|\partial_x^2 A|^2$ for perpendicular stripes, eq.~(\ref{Ampli_A_nonres_functional}). 
Thus, spatial variations of the amplitude $A({\bf r},t)$ affect the two functionals differently. 
The slow spatial variation of the control parameter $B_0(x)$ in $x$-direction is reflected in a spatial variation of the amplitude $A({\bf r},t)$. This increases both functionals. 
However, due to the different orders of $x$-derivatives, the functional for perpendicular stripes has a lower value \cite{Malomed:93.2,CrossHo}. 
Therefore, perpendicular stripes will be preferred for large $\delta_x$.

According to this reasoning, we predict stripes aligned perpendicular to the supercritical border for a large drop width $\delta_x$ and parallel for small $\delta_x$.
Therefore, we expect an orientational transition for medium values of $\delta_x$. 
Note, for these considerations only the contributions  $B_0$, $B_1$ and $B_2$ 
to the decomposition in eq.~(\ref{Eps_decomp}) 
are taken into account.
However,  the predicted orientational transition of stripes is rather  insensitive 
to these approximations  as confirmed  by simulations of the Brusselator in the next section.

\section{Numerical results for the Brusselator\label{results}}
\begin{figure}
\begin{center} \includegraphics[width=8cm]{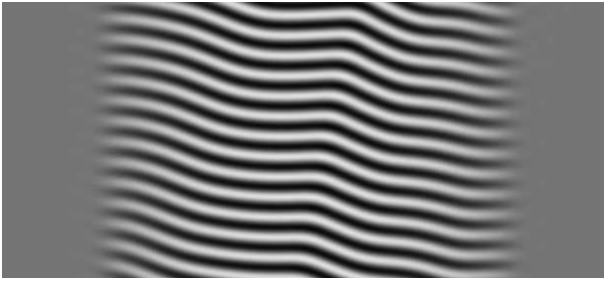} \end{center}
\caption{
Stripes favour a perpendicular orientation
with respect to slow control parameter drops  ($\delta_x = 5\lambda_c$).
Simulation of Brusselator started at $\beta_m=0.001$ and was slowly increased 
 to $\beta_m=0.05$. Parameters: $l_x=l_y=50\lambda_c$, $N_x=N_y=1024$.
Note: Only a cutout of the simulation is shown.
\label{Abb_perpstripes}}
\vspace{-2mm}
\end{figure}
In the previous part we found an orientational transition of stripe patterns by changing the width of control parameter drops.
This prediction is based on a reasoning including  approximations. 
Therefore, the effect is verified by simulations of the  Brusselator model,  cf. eqs.~(\ref{Brusselator}),
with  supercritical subdomains of width $L=20\lambda_c$, embedded in 
larger subcritical domains with overall system sizes $l_{x,y}$.
The model  is solved using a common pseudospectral method with periodic boundary conditions \cite{Kopriva:2009}
and $N_{x,y}$ modes, respectively.
We choose $\beta_0=-0.05$ and perturb the basic solution by small amplitude random noise.

\begin{figure}
\begin{center} 
\includegraphics[width=\columnwidth]{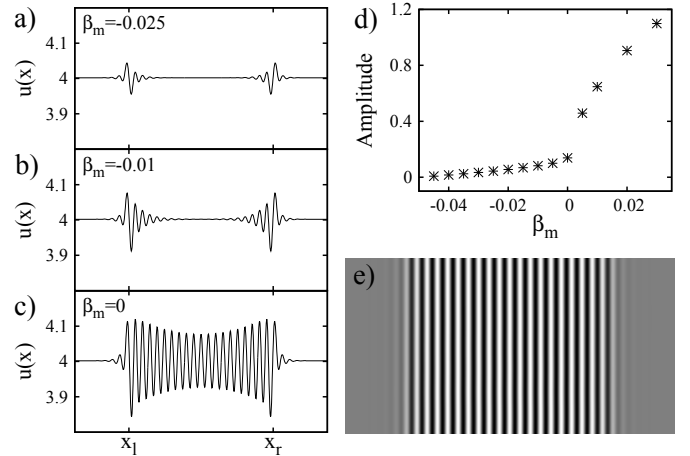} \end{center}
\vspace{-3mm}
\caption{Simulations of the Brusselator model with a narrow control parameter drop ($\delta_x=0.5\lambda_c$).
Cross sections of the two-dimensional stripe pattern for a) $\beta_m=-0.025$, b) $\beta_m=-0.01$, c) $\beta_m=0$.
d) The stripe amplitude as a function of $\beta_m$ implies an imperfect bifurcation.
e) Snapshot of the parallel stripes for $\beta_m=0.02$.
Simulation parameters: $l_x=50\lambda_c$, $l_y=25\lambda_c$, $N_x=1024$, $N_y=512$.
\label{Abb_parallelstripes}}
\end{figure}
For large widths $\delta_x$ of control parameter drops, {\it i.e.} slow variations of the control parameter, the preferred orientation of  a stripe pattern is nearly 
perpendicular to the borders of the supercritical domain, {\it i.e.} ${\bf q} \sim (0,q_c)$,
as shown in  fig.~\ref{Abb_perpstripes} for $\delta_x=5\lambda_c$.
This confirms the prediction in terms  of the amplitude equations in the
previous section (for bulk effects see Ref. \cite{Freund:2011.1}).
Similar  orientations are obtained for drop widths down to  about $\delta_x \simeq \lambda_c$.

For small $\delta_x$, {\it e.g}. $\delta_x=0.5\lambda_c$,
the stripes align parallel to the borders of the supercritical range, {\it i.e.} ${\bf q}_c\sim(q_c,0)$, as in fig.~\ref{Abb_parallelstripes}e) for $\beta_m=0.02$.
Moreover, localised Turing stripe patterns of finite amplitude  occur around the borders at $x_{l,r}$
already  at subcritical values of $\beta_m$ 
[see cross sections in fig.~\ref{Abb_parallelstripes}a) and b)].
For increasing $\beta_m$, they expand into the whole supercritical domain.
At the bulk threshold $\beta_m=0$ [fig.~\ref{Abb_parallelstripes}c)]
the stripes already have a finite amplitude throughout the range $[x_l,x_r]$. 
The maximum stripe amplitude of the stationary solution as a function of $\beta_m$ is shown in the bifurcation diagram in fig.~\ref{Abb_parallelstripes}d).
The form of the bifurcation is imperfect, as expected from the analysis on the basis of the amplitude equations in the previous section.

The two different  preferred stripe orientations  for large $ \delta_x=5\lambda_c$ in fig.~\ref{Abb_perpstripes} and 
small $\delta_x=0.5\lambda_c$ in fig.~\ref{Abb_parallelstripes}  clearly confirm an orientational transition of 
stripes in the supercritical domain depending on the width of the control parameter drop along its border.

We can further restrict the domain size by varying the control parameter simultaneously along the $x$- and $y$-direction, cf. eq.~(\ref{controldrop_2d}). 
In these rectangular domains \cite{Param1}, one can combine different drop widths $\delta_x$ and $\delta_y$ to trigger different stripe 
orientations as shown by four examples in fig.~\ref{Abb_rectangularboxes}. 
Combining, \textit{e.g.}, large drop widths at the long side of the rectangle with small drop widths at the short side
creates a remarkably uniform stripe pattern, cf. fig.~\ref{Abb_rectangularboxes}d).
Using different combinations of $\delta_{x,y}$ may be a promising tool for designing 
Turing patterns in localised light sensitive chemical reactions \cite{Steinbock:95.1}.

\section{Orientational transition regime}
\begin{figure}[functional]
 \includegraphics[width=\columnwidth]{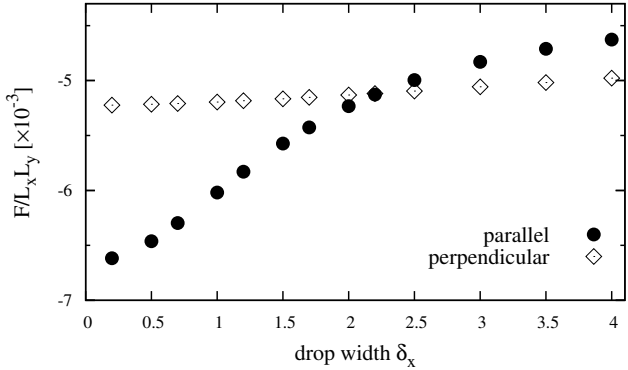}
 \caption{Comparison of the functional for stripes as function 
 of the drop width $\delta_x$ with the stripe wave vector ${\bf q}_c=(0,q_c)$  (filled circles)
 and ${\bf q}_c=(q_c,0)$ (open diamonds). Parameters: $\beta_0=-0.05$, $M=0.1$.
 \label{fig_func}}
\end{figure}

The orientational transition of stripes is deduced in terms of  amplitude equations
and confirmed by numerical simulations of the Brusselator model. 
The amplitude equations can be derived from the functionals, eqs.~(\ref{Ampli_A_resonance_functional}) and (\ref{Ampli_A_nonres_functional}). 
Calculating these functionals as a function of the drop width allows to determine the preferred orientation for this $\delta_x$. 
In the range where $F_\perp<F_\parallel$, a perpendicular stripe orientation is expected and vice versa.
For this purpose,  
we perform simulations of the amplitude equations for the two stripe orientations using the 
aforementioned pseudospectral algorithm (simulation parameters: $l_x=l_y=50\lambda_c$, $N_x=N_y=1024$, $L=20\lambda_c$,
$\beta_0=-0.05$, $\beta_m=0.05$). 
When the solutions reach the stationary state, the functionals displayed in fig.~\ref{fig_func}
are calculated.

The functional corresponding to perpendicular stripes 
in eq.~(\ref{Ampli_A_nonres_functional}) does not contain the non-adiabatic contributions
$B_1$ and $B_2$ to  the control parameter drop. 
Regardless of the assumptions
made for the justification of eq.~(\ref{Ampli_A_nonres_functional}) and the related 
amplitude equation, one may use
$\beta(x,\delta_x)$  instead of $B_0(x)$.
The functional then deviates only slightly from its constant value in the case of $B_0(x)$.
In addition, fig.~\ref{fig_func} shows that the  functional
with $\beta(x,\delta_x)$ is  nearly independent of $\delta_x$, {\it i.e.} stripes perpendicular to the border of the supercritical range are rather insensitive to the width $\delta_x$.

For parallel stripes, ${\bf q}=(q_c,0)$, the resonance effects covered by $B_1$ (and $B_2$)
are relevant and the associated functional is given in eq.~(\ref{Ampli_A_resonance_functional}).
The two functionals for the two different stripe orientations are shown as a function of the drop width $\delta_x$ in  fig.~\ref{fig_func}.
For narrow control parameter drops, \textit{i.e.} $\delta_x$ small, the functional for parallel stripes is significantly lower. 
Thus, the parallel orientation is preferred. 
However, the functional for parallel stripes strongly increases as a function of the drop width. 
The orientational transition takes place at the intersection of the two functionals. 
For larger $\delta_x$, the perpendicular orientation of the stripes is preferred.

\section{Summary and conclusions}
In this work, we identified and investigated a new class of finite pattern forming systems confined
by control parameter drops from super- to subcritical values.
These orient stripe patterns even
without boundary conditions for the relevant fields.
The stripe orientation depends on the width 
of the control parameter drops. 
We found a novel {\it orientational transition of stripe patterns} 
with respect to the 
borders as a function of the width  of  control parameter drops.

In light sensitive chemical  reaction-diffusion systems showing Turing patterns \cite{Epstein:1999.1,Epstein:2001.1} 
the transition length between the patterns (supercritical) and the homogeneous state (subcritical) may be varied by the length of a 
smooth  transition between illuminated and  dark areas.

The Swift-Hohenberg (SH) model \cite{Hohenberg:77.1}
is besides the Brusselator a further
 paradigmatic model for studying the formation of spatially periodic patterns \cite{CrossHo,Cross:2009}.
 It behaves differently
 with respect to control parameter drops along the border of a supercritical domain.
 The basic state of the Brusselator is a function of the control parameter $b$, cf. eq.~(\ref{Brusshomfix}).
 Control parameter drops thus also change the basic state of the bifurcation to Turing patterns.
 In contrast, the basic state $u_h=0$ of the SH model remains unchanged for spatially varying control parameters.
 The same applies to the mean-field model for block copolymers (see {\it i.e.} \cite{Weith:2013.1}).
 Therefore, we do not find the aforementioned orientational transition of stripe patterns in the SH or
 the block copolymer model.
 However, in  common systems where the basic state is also changed
 by control parameter variations, orientational transitions of stripe
 pattern are very likely.

Our results for stationary patterns may also be important
for traveling waves that occur for instance  in the cell biological MinE/MinD protein reaction 
on flat substrates  \cite{Schwille:2008.1,Schwille:2012.1}.
To mimic the effects of cell confinement in such extended experiments, 
reactive membranes were created in subdomains of the substrate \cite{Schwille:2012.1,Bassereau:2015.1}.
In this way, the traveling waves are restricted to the range above the functionalised parts of the membrane.
These may be interpreted as subdomains with a supercritical control parameter.
In this experiment the traveling waves align perpendicular to the borders of the functionalised area \cite{Schwille:2012.1}.
It is very likely that this orientational behaviour is again governed by generic principles 
as discussed in this work and
specific molecular reaction schemes or three-dimensional effects provide quantitative modifications
\cite{Schwille:2012.1,Halatek:2012.1,Halatek:2014.1}.
Is the complex behavior of MinE/MinD oscillations in further restricted domains, as investigated recently  in  Ref.~\cite{DekkerC:2015.1}, determined by the specific properties of the kinetic reaction models? 
Or do again generic principles of pattern formation play a leading role as described in this work?

\acknowledgments
Enlightening discussions with M. Hilt and M. Wei\ss \ are
gratefully  acknowledged.


\end{document}